%% file: arXiv.tex
\documentclass[prl,reprint,superscriptaddress]{revtex4-2}

\input{preamble}

\makeatletter
\def\l@subsubsection#1#2{}
\makeatother

\usepackage{color}

\begin{document}

\title{Efficient Preparation of Quantum States via Randomized Truncation}

\author{Yue Wang}
\affiliation{QICI Quantum Information and Computation Initiative, School of Computing and Data Science,
The University of Hong Kong, Pokfulam Road, Hong Kong SAR, China}
\affiliation{Shenzhen International Quantum Research Institute, Shenzhen 518000, Guangdong, China}

\author{Xiao-Ming Zhang}
\email{phyxmz@gmail.com}
\affiliation{Key Laboratory of Atomic and Subatomic Structure and Quantum Control (Ministry of Education), Guangdong Basic Research Center of Excellence for Structure and Fundamental Interactions of Matter, School of Physics, South China Normal University, Guangzhou 510006, China}
\affiliation{Guangdong Provincial Key Laboratory of Quantum Engineering and Quantum Materials, Guangdong-Hong Kong Joint Laboratory of Quantum Matter, Frontier Research Institute for Physics, South China Normal University, Guangzhou 510006, China}

\author{Xiao Yuan}
\email{xiaoyuan@pku.edu.cn}
\affiliation{Center on Frontiers of Computing Studies, Peking University, Beijing 100871, China}
\affiliation{School of Computer Science, Peking University, Beijing 100871, China}

\author{Qi Zhao}
\email{zhaoqi@cs.hku.hk}
\affiliation{QICI Quantum Information and Computation Initiative, School of Computing and Data Science,
The University of Hong Kong, Pokfulam Road, Hong Kong SAR, China}

\begin{abstract}
While the preparation of a general quantum state is challenging, realistic problem instances—such as those encountered in quantum chemistry and quantum machine learning—typically exhibit hierarchical amplitude structures, consisting of a small number of large components alongside a vast number of small but non-negligible ones.
Standard approaches deterministically truncate the small amplitude would incur an approximation error that scales linearly with the discarded amplitude mass, enforcing a rigid trade-off between precision and circuit depth. 
Here, we  circumvent the challenge by introducing a randomized state-preparation protocol with probabilistic amplification of small amplitudes using ensembles of low-complexity circuits. 
Analytically, we prove that this approach significantly reduces the number of encoded amplitudes—halving the requirement for exponentially decaying states and offering asymptotically larger gains for heavy-tailed power-law decays. 
Numerical simulations on LiH molecular wavefunctions and deep-learning-derived states demonstrate reductions of up to 99\% in CNOT and T-gate counts compared with deterministic methods. These results establish a resource-efficient paradigm for initializing complex states, relaxing gate-synthesis precision requirements for both near-term and fault-tolerant hardware, and improving the end-to-end feasibility of quantum computing.
\end{abstract}
\maketitle

\prlsection{Introduction}Preparing a given quantum state is a foundational primitive in quantum computing and frequently the dominant bottleneck in end-to-end algorithms. Applications ranging from quantum simulation of molecular and materials systems~\cite{linNearoptimalGroundState2020,aspuru2005simulated,mcardle2020quantum,reiher2017elucidating,mcardle2019variational} and quantum algorithms for differential equations~\cite{berryQuantumAlgorithmLinear2017,liuEfficientQuantumAlgorithm2021,an_quantum_2022,anTheoryQuantumDifferential2023,kroviImprovedQuantumAlgorithms2023,fangTimemarchingBasedQuantum2023} to quantum machine learning pipelines~\cite{biamonte2017quantum,havlivcek2019supervised,lloyd2014quantum} typically assume access to a highly structured initial state. {Specifically, the target state may exhibit a pronounced amplitude hierarchy: a compact core of dominant configurations accompanied by a heavy tail of many small but non-negligible components, reflecting underlying physical structures such as many-body correlation.} The cost of preparing such states can rival or exceed the subsequent quantum processing~\cite{aaronson2015read}; nevertheless, state preparation is often idealized as a black-box oracle and omitted from complexity analyses. When preparation dominates, the putative algorithmic speedups can remain inaccessible on hardware.

Preparing an arbitrary $n$-qubit state requires resources exponential in $n$~\cite{PleschQuantumStatePreparation2011,ShendeSmallerTwoQubit2004}. For structured targets, sparse-state preparation protocols~\cite{gleinigEfficientAlgorithmSparse2021a,zhangQuantumStatePreparation2022,deverasDoubleSparseQuantum2022,maoOptimalCircuitSize2024,luoCircuitComplexitySparse2024} substantially reduce overhead, but the cost still scales linearly with the number of encoded amplitudes $d$, with a circuit-size lower bound $\Omega(d+n)$. {As a result, achieving high fidelity for hierarchical states requires retaining and coherently encoding many small-amplitude components from the heavy tail, which rapidly increases circuit size and depth.} This becomes prohibitive for targets with millions of relevant components, such as configuration-interaction wavefunctions in quantum chemistry or dense superpositions in data-driven tasks. Consequently, practical implementations rely on amplitude truncation: discarding coefficients below a threshold to reduce circuit complexity.

Deterministic truncation, however, suffers from a severe linear cost--accuracy trade-off. To achieve trace-distance error at most $\varepsilon$, one must retain enough amplitudes so that the discarded $\ell_2$ weight is $\mathcal{O}(\varepsilon)$. Small-magnitude amplitudes impose a double resource penalty: they increase the total number of retained components $d$ while simultaneously demanding higher gate-synthesis precision, leading to substantial fault-tolerant $T$-gate overhead~\cite{mooneyCostoptimalSinglequbitGate2021,rossOptimalAncillafreeClifford+T2016}. In addition, components at large Hamming distance from the dominant core can incur high CNOT costs due to high-control-number rotations. These compounding demands motivate approximation strategies that move beyond the rigid limits of deterministic truncation and instead exploit the hierarchical structure intrinsic to realistic states.

Here we introduce a randomized state-preparation protocol that fundamentally improves this trade-off. Rather than deterministically discarding the heavy tail, we approximate the target using a randomized ensemble of low-complexity circuits: each circuit instance retains all large-magnitude amplitudes and selectively amplifies a small-magnitude component. Sampling from this ensemble with probabilities matched to the amplified coefficients produces a mixed state that recovers the target in expectation while keeping individual circuit costs low. Importantly, amplifying small coefficients also reduces fault-tolerant resource requirements by lifting those contributions out of the highest-precision regime. We prove that the trace-distance error of the resulting randomized ensemble scales quadratically with the discarded $\ell_2$ weight, i.e., $\mathcal{O}(\varepsilon^2)$ for discarded weight $\mathcal{O}(\varepsilon)$, thereby breaking the linear scaling inherent to conventional truncation. This yields a halving of the number of encoded amplitudes for exponentially decaying states and even greater savings for power-law decay, reducing both gate count and circuit depth.

We validate the protocol on realistic quantum chemistry instances (LiH) and on synthetic power-law, many-body (transverse-field Ising), and machine-learning (ResNet) states. For a 10-qubit synthetic power-law state, we observe reductions of up to $98.9\%$ in CNOT counts and $98.5\%$ in $T$-gate counts at fixed accuracy. The method is compatible with near-term and fault-tolerant regimes: it replaces a single deep monolithic circuit with sampling over simpler circuits, easing the state-preparation bottleneck and expanding the class of quantum states accessible on hardware.

\prlsection{Randomized Framework}Given a general $n$-qubit quantum state 
$\ket{\psi} = \sum_{i=0}^{2^n-1} \alpha_i \ket{i}$,
we consider constructing an approximated state $\rho_{\text{approx}}$ (typically from $\ket{0}^{\otimes n}$) such that the error is below a threshold, i.e., $\| |\psi\rangle\langle \psi|-\rho_{\text{approx}} \|_1\leqslant\varepsilon$. Suppose $\ket{\psi}$ has $d$ nonzero amplitudes $\alpha_i$ and no error is allowed; then any state-preparation protocol has a circuit-size lower bound $\Omega(d+n)$ \footnote{This is a trivial lower bound on the cost of state preparation; the state-of-the-art protocol achieves $\bigO{dn/\log(dn) + n}$ \cite{luoCircuitComplexitySparse2024}}. This cost can be prohibitive in practice for large $d$. Amplitude truncation discards insignificant coefficients, thereby reducing $d$ at a cost of  introducing an approximation error. Such truncation methods are widely used, including in low-rank approximations for quantum machine learning~\cite{biamonte2017quantum, lloydQuantumPrincipalComponent2014, liuProvablyEfficientQuantum2024}, finite-range interaction truncation in condensed-matter simulations~\cite{babbushLowDepthQuantum2017, babbushEncodingElectronicSpectra2018}, and quantum chemistry~\cite{lanyonQuantumChemistryQuantum2010,mcardleQuantumComputationalChemistry2020,romeroStrategiesQuantumComputing2018,helgaker2013molecular}.

We partition indices into significant ($A=\{i:|\alpha_i|\ge t\}$) and negligible ($B=\{i:|\alpha_i|<t\}$) sets according to a cutoff threshold $t$.
Discarding set $B$ and renormalizing the state as $\ket{\psi_A} / \|\ket{\psi_A}\|$, where $ \ket{\psi_A} := \sum_{i \in A} \alpha_i \ket{i}$, 
then the cutoff error is bounded by $\bigO{\sqrt{\sum_{i \in B} |\alpha_i|^2}}$.
The choice of partitioning threshold depends on the specific task, provided that the truncated amplitudes carries sufficient weight to reduce complexity within the target error. In quantum chemistry, for instance, set $A$ might contain the Hartree--Fock (HF) coefficients, while set $B$ includes higher-order excitations. 

Building on the $A/B$ partition introduced above, we now ask whether one can retain the influence of the discarded tail $B$ without paying the full cost of encoding all its amplitudes in a single circuit.  

To resolve the problem, we apply randomization, which has emerged as a powerful tool for improving many aspects in quantum computing~\cite{wallman2016noise,childsFasterQuantumSimulation2019,campbellRandomCompilerFast2019a,wanRandomizedQuantumAlgorithm2022,huang2020predicting,hastingsTurningGateSynthesis2016,campbellShorterGateSequences2017,wangFasterQuantumAlgorithms2024a,martynHalvingCostQuantum2024}, to the 
state-preparation problem.  Our strategy is to avoid encoding the entire small-amplitude tail coherently in a single circuit. Instead, we use classical randomness to distribute the tail amplitudes across an ensemble of simpler states: each member of the ensemble retains the large-amplitude structure in $A$ but only includes an amplified subset of $B$, chosen with carefully designed probabilities. Each instance is only slightly more expensive than preparing the truncated state $\ket{\psi_A}$, yet the resulting mixed state recovers the target with quadratically smaller trace-distance error. The improved accuracy-cost trade-off implies the randomized approach uses smaller quantum circuit to achieve a certain cutoff error level. 

We illustrate our framework in Alg.~\ref{alg:randomized_state_prep}. Each state $\ket{\psi_m}$ in the ensemble contains all amplitudes in $A$ and a single element in $B$ with an amplified coefficient. We choose normalized probabilities so that the reconstruction condition holds:
$\sum_{m} p_m \frac{\alpha_m}{p_m}    \ket{m} = \sum_{i \in B} \alpha_i \ket{i},$ where $p_m = |\alpha_m| \big/ \sum_{i \in B} |\alpha_i|$ is the selection probability corresponding to index $m$.  By carefully accounting for the contribution of each term to the final error, our main result is as follows.

\begin{algorithm}[t]
    \caption{Randomized Quantum State Preparation Protocol}
    \label{alg:randomized_state_prep}
    
    \KwIn{Quantum state $\ket{\psi} = \sum_i \alpha_i \ket{i}$ with indices partitioned into $A = \{ i: |\alpha_i| \geq t \}$ and $ B = \{ i : |\alpha_i| < t \}$.}

    \KwOut{Approximated quantum state $\rho_{\text{approx}}$ as a randomized ensemble of states $\ket{\tilde\psi_m}$.}

    Compute sampling probabilities $p_m = \frac{|\alpha_m|}{\sum_{i \in B} |\alpha_i|}$ for $m \in B$.
    
    Define the corresponding normalized state:
    \begin{equation}
    \ket{\tilde\psi_m} = \Gamma^{-1}\left(\sum_{i \in A} \alpha_i \ket{i} + \frac{\alpha_m}{p_m} \ket{m}\right),
    \end{equation}
    where $\Gamma = \sqrt{\sum_{i \in A} |\alpha_i|^2 + \frac{|\alpha_m|^2}{p_m^2}}$ is a  normalization factor independent of $m$.

    Sample an index $m$ with probability $p_m$ and prepare $\ket{\tilde{\psi}_m}$. The final prepared state is a mixture:
    \begin{equation}
    \rho_{\text{approx}} = \sum_{m} p_m \ket{\tilde{\psi}_m} \bra{\tilde{\psi}_m}.
    \end{equation}
\end{algorithm}


\begin{theorem}
\label{thm: error_bound}
(Informal) Let $\ket{\psi}=\ket{\psi_A}+\sum_{i \in B}\alpha_i\ket{i}$, and $\epsilon^2:=\sum_{i \in B}|\alpha_i|^2$,
with real amplitudes $\alpha_i$.  Assume $S:=\sum_{i \in B}|\alpha_i|=\bigO{\epsilon}.$ The output state of Alg.~\ref{alg:randomized_state_prep} satisfies
\begin{equation}
    \|\rho_{\mathrm{approx}}-\ket{\psi}\bra{\psi}\|_1\le a^2+2b=\bigO{\epsilon^2}.
\end{equation}
\end{theorem}

Theorem \ref{thm: error_bound} (proof in Appendix) indicates that our protocol leads to a quadratically improved error scaling. {The same truncated tail weight that would induce an $\mathcal{O}(\epsilon)$ state error under deterministic truncation only contributes $\mathcal{O}(\epsilon^2)$ here, allowing substantially fewer encoded amplitudes at a fixed target fidelity.} The assumption of $S=\bigO{\epsilon}$ is justified in Appendix. Briefly, it holds if the coefficients with indices in $B$ decay as a power law with exponent greater than 1, or decay exponentially. The theorem holds even if some $p_m$ are exponentially small, and create exponential amplifications. Nevertheless, in that regime, the corresponding $\alpha_m$ and the selection probability are also exponentially small ensuring that the overall cutoff error remains bounded by $\bigO{\epsilon^2}$.

\prlsection{Resource savings}Theorem~\ref{thm: error_bound} demonstrates that randomization suppresses the approximation error from $\mathcal{O}(\epsilon)$ to $\mathcal{O}(\epsilon^2)$. The quadratic suppression directly reduces the resources required to reach a prescribed preparation error, thereby lowering the overall resource cost in practical end-to-end implementations. To quantify the resource advantage, we compare the number of retained amplitudes, $K$, required to achieve a fixed predetermined trace-distance error $\hat{\epsilon}$. Since circuit depth typically scales linearly with $K$, the ratio of required amplitudes for deterministic ($K_{\det}$) versus randomized ($K_{\rand}$) truncation quantifies the relative efficiency. We evaluate this scaling under two standard decay models.

\begin{corollary}
\label{cor:exp_decay}
For states with exponentially decaying amplitudes $|\alpha|_{(j)} \le C r_{\Exp}^{j}$ ($0<r_{\Exp}<1$), the randomized protocol requires half the amplitudes of the deterministic approach, satisfying the asymptotic relation $K_{\det} = 2 K_{\rand}$.
\end{corollary}

The reduction stems from the logarithmic dependence of $K$ on the target error. For exponential decay, the tail weight scales as $\epsilon(K) \sim r_{\Exp}^K$. Achieving the target error $\hat{\epsilon}$ deterministically requires $K_{\det} \propto \log(1/\hat{\epsilon})$. In contrast, the randomized error scales as $\epsilon(K_{\rand})^2 \sim r_{\Exp}^{2K_{\rand}}$. Matching this to $\hat{\epsilon}$ yields $K_{\rand} \propto \frac{1}{2}\log(1/\hat{\epsilon})$, doubling the effective convergence rate per encoded amplitude. This exponential decay phenomenon exists prevalently in nature, as seen in CI states and the thermal states of gapped Hamiltonians.

\begin{corollary}
\label{cor:power_decay}
For states with power-law decay $|\alpha|_{(j)} \le C j^{-r_{\pow}}$ ($r_{\pow}>1$), the resource savings scale as
\begin{equation}
\frac{K_{\det}}{K_{\rand}} = \mathcal{O}\left(\hat{\epsilon}^{-\tfrac{1}{2r_{\pow}-1}}\right).
\end{equation}
\end{corollary}

Under power-law decay, the truncation error diminishes algebraically as $\epsilon(K) \sim K^{-(r_{\pow}-1/2)}$. Consequently, achieving $\hat{\epsilon}$ implies $K_{\det} \sim \hat{\epsilon}^{-\frac{2}{2r_{\pow}-1}}$, whereas the quadratic error suppression of the randomized protocol yields $K_{\rand} \sim \hat{\epsilon}^{-\frac{1}{2r_{\pow}-1}}$. The ratio $K_{\det}/K_{\rand}$ diverges as $\hat{\epsilon} \to 0$, indicating that for heavy-tailed distributions, which are common in many-body physics and data loading, randomization offers a $o(\hat\epsilon)$ cost advantage that grows with the desired precision.

Beyond amplitude counts, this early truncation yields significant gate-level savings. In fault-tolerant compilations, the smallest tail amplitudes typically require the longest sequences of $T$ gates to resolve. By discarding these terms without sacrificing accuracy, the randomized protocol eliminates the most computationally expensive components of the circuit.

\prlsection{Numerical Results}The randomized approach is particularly beneficial for quantum chemistry~\cite{lanyonQuantumChemistryQuantum2010,mcardleQuantumComputationalChemistry2020,romeroStrategiesQuantumComputing2018,helgaker2013molecular}, notably in configuration interaction (CI) methods. In CI, the many-electron wavefunction is represented as a linear combination of Slater determinants, inherently exhibiting an amplitude-decay structure. Typically, the largest amplitudes correspond to the HF determinant and low-lying excitations, whereas amplitudes for higher-order excitations diminish rapidly due to large energy differences in the denominators.

Preparing the full configuration interaction (FCI) state is computationally prohibitive due to the exponential growth in determinants with electron and orbital counts. Consequently, practical workflows employ truncated expansions. Such truncations integrate naturally with our randomized preparation, reducing quantum resources by avoiding explicit reconstruction of the entire FCI state in a single instance.

We validate our framework on the ground state of LiH and a synthetic 10-qubit power-law state whose sorted amplitudes follow $|\alpha|_{(j)} \propto j^{-5}$ with random sign. Additionally, to demonstrate broad applicability, we evaluate the ground state preparation of transverse-field Ising model (TFIM; $N=11$, $J=h=1$) and a 10-qubit state constructed from parameters of a trained ResNet~\cite{he2016deep}. Although the data structures in TFIM and ResNet do not strictly follow the specified power law or geometric decay, we still observed a reduction in the cutoff error at the same cutting threshold as shown in Fig.~\ref {fig:additional}. These represent both physically structured states and high-dimensional data-driven states, demonstrating the broad applicability of this randomized protocol.

\begin{figure*}[t]
    \centering
    \includegraphics[width=.9\linewidth]{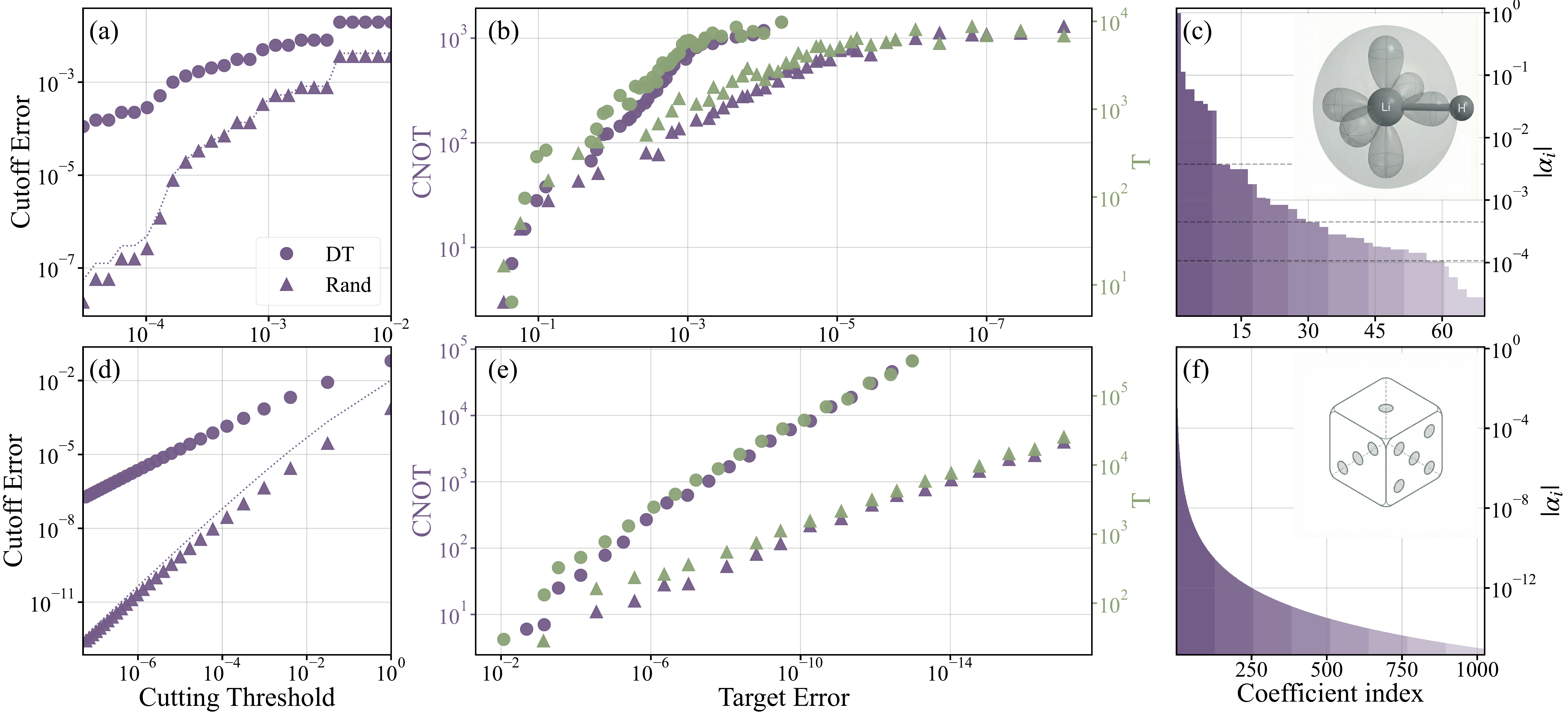}
    \caption{Comparison of randomized and deterministic truncation protocols for molecular and power-law states. (a, d) Cutoff error versus cutting threshold for (a) the LiH ground state and (d) a synthetic 10-qubit power-law state ($r_{\text{pow}} = 5$). Circles and triangles denote deterministic and randomized methods, respectively. Dotted lines indicate theoretical upper bounds. (b, e) Resource cost scaling for (b) LiH and (e) the power-law state. The plots display the CNOT (green, left axis) and $T$-gate (purple, right axis) counts against the target error. The randomized protocol yields significant resource reductions: up to 82.2\% (81.2\%) for CNOT ($T$-gate) in LiH, and 98.9\% (98.5\%) for the power-law state. (c, f) Amplitude decay spectra for (c) the LiH CI state and (f) the power-law state.}
    \label{fig:main}
\end{figure*}

For LiH, we map the ground state wavefunction of LiH to 12 qubits
and form the ensemble density matrix by truncating CI coefficients. We employ the Gleinig-Hoefler protocol~\cite{gleinigEfficientAlgorithmSparse2021a} to generate the quantum circuit that prepares the desired quantum state. While we utilize this specific protocol as a representative baseline, the resource savings in our approach derive fundamentally from reducing the number of encoded amplitudes; consequently, we expect similar advantages to extend to other state-preparation algorithms. We further decompose the generated state into Clifford$+$single qubit SU(2) rotation gates. Figs.~\ref{fig:main}(a) and (b) show that we achieve low cutoff error relative to the FCI state at substantially reduced gate cost compared to the deterministic truncation method. Specifically, for the chemistry instance, we achieve a maximum CNOT reduction of $82.2\%$ (reducing from 962 to 171 CNOTs) at a trace-norm error of $5.86 \times 10^{-4}$. To probe the power-law regime, we compute the quantum circuit for the synthetic 10-qubit state ($r = 5$) at different truncation level, shown in Figs.~\ref{fig:main}(d) and (e). Comparing deterministic truncation to our randomized mixture at matched cutoff error, the resource savings are even more pronounced, where we observe a maximum CNOT reduction of $98.9\%$ (reducing from 66,607 to 742 CNOTs) at a trace-norm error of $1.04 \times 10^{-13}$. 

We also analyze the $T$-gate count through ancillary-free rotation gate synthesis~\cite{rossOptimalAncillafreeClifford+T2016}. We decompose rotation gates into sequences of $T$ gates using the upper-bound $\approx 3\log_2(1/\delta)$, where $\delta$ is the precision. To maintain small gate-synthesis error, we set the per-gate synthesis precision to $\delta = \theta_{\min}/m$, where $m$ is the number of rotation gates and $\theta_{\min}$ is the minimum rotation angle.  As shown in Figs.~\ref{fig:main} (b) and (e), the randomized protocol yields substantial $T$-gate savings. For the LiH CI state, we observe a maximum T-gate reduction of $81.20\%$ and for the power-law state, the reduction is $98.45\%$. 

These savings arise from two factors. First, the reduction in the number of encoded amplitudes, where we observe a $50\%$ reduction in the number of coefficients for the LiH instance and a $95\%$ reduction for the power-law instance. Second, amplifying small coefficients mitigates the need for extremely high-precision multi-control rotation gates. By amplifying the smallest coefficients, we effectively increase the minimum rotation angle $\theta_{\min}$, thereby reducing the $T$-gate depth required to resolve the state components. Moreover, these small tail amplitudes also require high control numbers, which decompose into a large number of CNOT gates. This explains why gate reductions exceed the amplitude count reduction. 

In the LiH instance, the ground state in the computational basis is highly irregular. Small-amplitude indices often appear as isolated outliers in the state-preparation tree structure, requiring deep branches of multi-controlled gates to address. This structural irregularity causes the non-monotonic scaling behavior observed in Fig.~\ref{fig:main}(b) and (e), where the removal of specific costly outliers yields additional savings.

\begin{figure}[t]
    \centering
    \includegraphics[width=\linewidth]{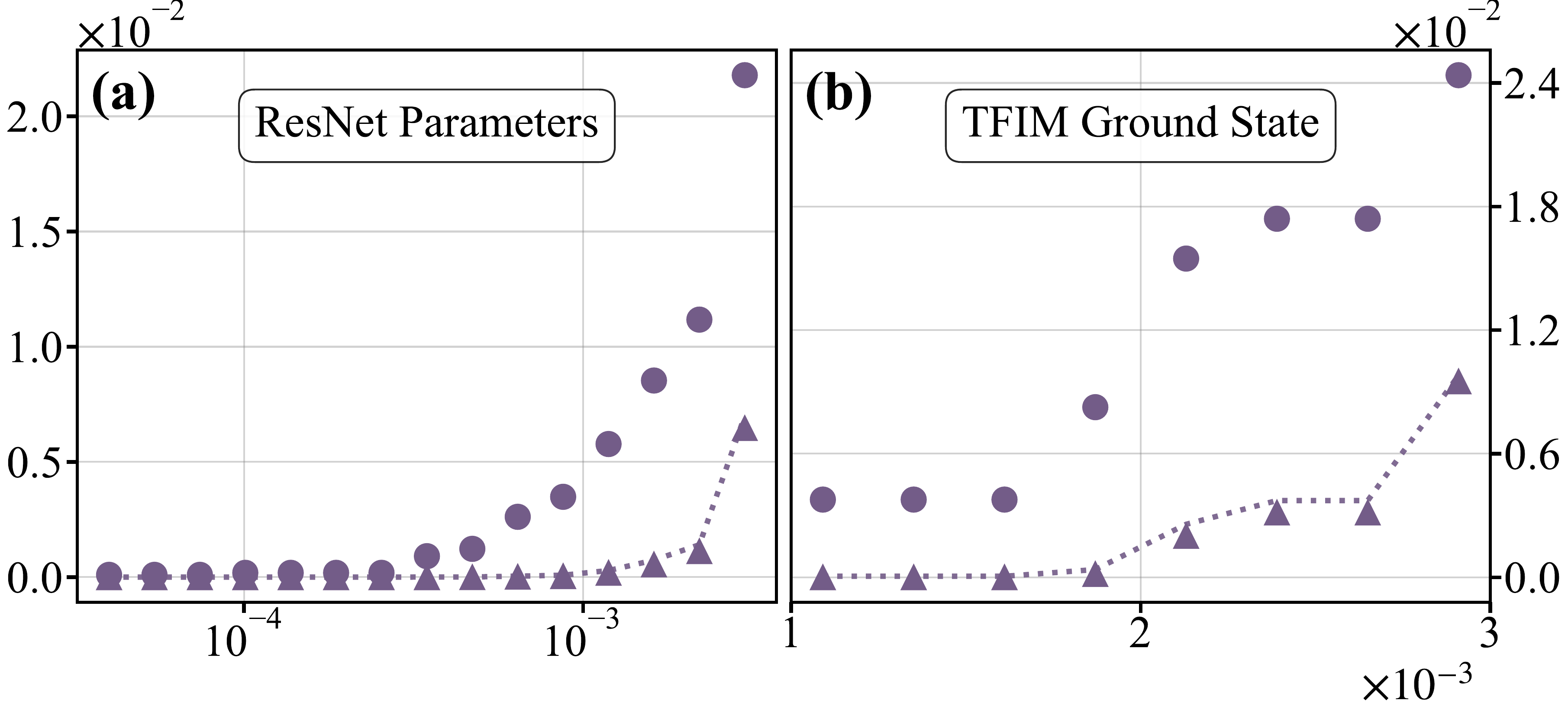}
    \caption{Performance on state preparation in machine learning and many-body physics. The plots display the trace-distance error (vertical axis) versus the cutting threshold (horizontal axis) for (a) a 10-qubit state encoding trained ResNet parameters~\cite{he2016deep} and (b) the ground state of the Transverse-field Ising model (TFIM; $N=11$, $J=h=1$). Circles represent the deterministic truncation method, while triangles denote the randomized protocol, illustrating a significant suppression of error. The dotted line indicates the theoretical upper bound for the randomized approximation.}
    \label{fig:additional}
\end{figure}

In Fig.~\ref{fig:additional}, we demonstrate the versatility of the method on a ResNet-derived state and the ground state preparation of Transverse-field Ising model (TFIM). For ResNet~\cite{he2016deep}, we construct a quantum state from the parameters of a trained ResNet neural network on CIFAR-10 image test set. We load the final parameter vector (526 floating-point values), normalize it, and map it to amplitudes of a 10-qubit state, filling the first 526 computational-basis states. Finally, we construct the ground state of the $11$-qubit TFIM. The Hamiltonian is $H = -J \sum_{i=0}^{N-1} Z_i Z_{i+1} - h \sum_{i=0}^{N-1} X_i$, where $X_i$ and $Z_i$ act on qubit $i$. The ground state is a dense superposition over all $2^{11}$ computational-basis states.  In both cases, the randomized protocol consistently achieves lower trace-norm error at fixed cost compared to deterministic truncation, confirming the method's applicability across physics and machine learning domains.


\prlsection{Discussion}We presented a randomized state preparation approach that reduces circuit complexity while maintaining low cutoff error. The method leverages hierarchical amplitude structures common in practice, such as chemistry, condensed-matter systems, and machine learning. The randomized approach avoids preparing all tiny amplitudes in a single circuit instance and thus reduces the required resources. Unlike traditional deterministic methods, which uniformly address all coefficients, our approach prioritizes amplitude subsets based on significance, achieving a practical balance between cutoff error and circuit complexity. {Implications are twofold. For near-term devices, sampling over simpler circuits reduces depth and makes state preparation compatible with coherence limits. For fault-tolerant architectures, the same mechanism lifts tiny amplitudes out of the highest-precision regime, reducing the synthesis overhead of small-angle rotations and the associated $T$-gate cost.} Our analysis establishes error bounds proportional to the square of the discarded tail weight, yielding a quadratic improvement over direct truncation. Numerical simulations demonstrates $\sim$99\% gate number reduction and several order of magnitudes of cutoff error improvements, corroborating theoretical gains in both accuracy and resource utilization.


In practice, one may amplify a subset S of size $M \ge 1$ in each instance. Distributing the amplification across multiple small coefficients reduces the per-instance amplification magnitude and correspondingly mitigates the diagonal (quadratic) contribution to the mixture error. To avoid unwanted cancellations that would lead to an unbounded amplification, each subset should be of the same sign.

Our framework is well suited to hardware implementation. The states in the ensemble $\ket{\tilde{\psi}_m}$ are very similar, differing only in the sign and the computational basis vector of the amplified term. Thus, the state-preparation procedures for the states in the ensemble are highly similar, facilitating practical implementation of the randomized scheme. In summary, our approach reduces error for truncated state preparation, making it promising for near-term and fault-tolerant devices. The rigorous guarantees ensure that prepared states closely approximate targets with controllable error, enabling more efficient implementations of algorithms that rely on complicated state preparation.

\prlsection{Note added}During the preparation of the manuscript, we note a recent work by Harrow et al.~\cite{harrowRandomizedTruncationQuantum2025}, who studied optimal randomized approximation of arbitrary pure states by optimizing convex mixtures of $k$-sparse states. While the studied problems are related, the two methods are different and complementary.

\begin{acknowledgments}
    Q.Z. acknowledges funding from Innovation Program for Quantum Science and Technology via Project 2024ZD0301900, National Natural Science Foundation of China (NSFC) via Project No. 12347104 and No. 12305030, Guangdong Basic and Applied Basic Research Foundation via Project 2023A1515012185, Hong Kong Research Grant Council (RGC) via No. 27300823, N\_HKU718/23, and R6010-23, Guangdong Provincial Quantum Science Strategic Initiative No. GDZX2303007, HKU Seed Fund for Basic Research for New Staff via Project 2201100596. 
\end{acknowledgments}

\bibliography{bibli}

\appendix
\onecolumngrid

\section{State-Level Mixing Lemma}
We first propose a state-level mixing lemma, where we consider the random mixing of quantum states. 
\begin{lemma}[State-Level Generalized Mixing Lemma]
\label{lem:mix}
Let $\ket{\psi}$ be a target pure quantum state, and let $\{\ket{\psi_l}\}$ be a family of pure states with associated probabilities $\{p_l\}$ ($\sum_l p_l = 1$). Assume the following bounds hold:
\begin{equation}
    \label{eq:assumptions}
    \|\ket{\psi_m} - \ket{\psi}\| \le a \quad \forall m, 
    \qquad 
    \left\| \sum_m p_m \ket{\psi_m} - \ket{\psi} \right\| \le b,
\end{equation}
for some constants $a, b > 0$, where $\|\cdot\|$ denotes the Euclidean norm. Define the mixed density matrices as $ \rho_{\mathrm{mix}} := \sum_l p_l \ket{\psi_l}\bra{\psi_l}$.

Then, the trace norm between the mixed and target states satisfies
\begin{equation}
    \label{eq:trace_bound}
    \|\rho_{\mathrm{mix}} - \ket{\psi}\bra{\psi}\|_1 \le a^2 + 2b.
\end{equation}
\end{lemma}

\begin{proof}
Define the deviation vectors $\delta_m := \ket{\psi_m} - \ket{\psi}$, which satisfy $\|\delta_m\| \le a$. Let their average deviation be
\begin{equation}
    \label{eq:delta_avg}
    \Delta := \sum_m p_m \delta_m = \sum_m p_m (\ket{\psi_m} - \ket{\psi}),
\end{equation}
which satisfies $\|\Delta\| \le b$ by assumption \eqref{eq:assumptions}.

Expanding each projector:
\begin{equation}
    \label{eq:proj_expansion}
    \ket{\psi_m}\bra{\psi_m} 
    = (\ket{\psi} + \delta_m)(\bra{\psi} + \delta_m^\dagger) 
    = \rho + \ket{\psi}\bra{\delta_m} + \ket{\delta_m}\bra{\psi} + \ket{\delta_m}\bra{\delta_m}.
\end{equation}
Taking the weighted sum and subtracting $\rho$, we obtain:
\begin{align}
    \rho_{\mathrm{mix}} - \rho 
    &= \sum_m p_m \left( \ket{\psi}\bra{\delta_m} + \ket{\delta_m}\bra{\psi} + \ket{\delta_m}\bra{\delta_m} \right) \notag \\
    &= \ket{\psi}\bra{\Delta} + \Delta\bra{\psi} + \sum_m p_m \ket{\delta_m}\bra{\delta_m}.
    \label{eq:rho_diff}
\end{align}

We now bound the trace norm of each term in Eq.~\eqref{eq:rho_diff}.

Define the operator
\begin{equation}
    \label{eq:X_def}
    X := \ket{\psi}\bra{\Delta} + \Delta\bra{\psi}.
\end{equation}
This is a Hermitian operator of rank at most 2. By the triangle inequality and the identity $\|\ket{u}\bra{v}\|_1 = \|u\|\|v\|$, we get
\begin{equation}
    \label{eq:X_bound}
    \|X\|_1 
    \le \|\ket{\psi}\bra{\Delta}\|_1 + \|\Delta\bra{\psi}\|_1 
    = 2 \|\psi\| \cdot \|\Delta\| = 2\|\Delta\| \le 2b.
\end{equation}

Each term $\ket{\delta_m}\bra{\delta_m}$ has trace norm $\|\delta_m\|^2 \le a^2$. Thus,
\begin{equation}
    \label{eq:quad_term}
    \left\| \sum_m p_m \ket{\delta_m}\bra{\delta_m} \right\|_1 
    = \sum_m p_m \|\delta_m\|^2 \le a^2 \sum_m p_m = a^2.
\end{equation}

Combining Eqs.~\eqref{eq:X_bound} and \eqref{eq:quad_term} in Eq.~\eqref{eq:rho_diff}, we obtain the desired bound:
\begin{equation}
    \|\rho_{\mathrm{mix}} - \rho\|_1 
    \le \|X\|_1 + \left\| \sum_m p_m \ket{\delta_m}\bra{\delta_m} \right\|_1 
    \le 2b + a^2.
\end{equation}
This proves the lemma.
\end{proof}

This lemma provides a bound on how well a convex mixture of approximating pure states reproduces the original pure state at the level of density operators. Importantly, the bound depends only on the worst-case deviation $a$ and the average deviation $b$. The probabilities $p_m$ may be arbitrarily small---the proof depends only on their normalization $\sum_m p_m = 1$. This lemma is particularly useful when designing randomized state preparation protocols where each sampled state is close to the target, but their average is even closer.

\section{Trace Norm Error Bound}
We can bound the error in the random mixing protocol by finding the bound on $a$ and $b$.  We apply Lemma~\ref{lem:mix} to our randomized protocol. Consider an $n$-qubit state $\ket{\psi} = \sum_{i=0}^{2^n-1} \alpha_i \ket{i}$. We partition the indices into a "kept" set $A = \{ i : |\alpha_i| \ge t \}$ and a "tail" set $B = \{ j : |\alpha_j| < t \}$. The state can be written as:
\begin{equation}
  \ket{\psi} = \ket{\psi_A} + \sum_{j\in B}\alpha_j\ket{j},
  \qquad
  \|\psi_A\|^2=1-\epsilon^2,
  \qquad
  \epsilon^2:=\sum_{j\in B}\alpha_j^2.
\end{equation}
We construct an ensemble $\{\ket{\tilde{\psi}_m}\}$ (normalized) with probabilities $p_m$.

\begin{theorem}[Random mixing error bound]
Assume
\begin{equation}
\label{eq:l1small}
   S := \sum_{j\in B}|\alpha_j| \le c\epsilon = \bigO{\epsilon}
\end{equation}
on the tail coefficients. For every $m\in B$, define the probability $p_m:=|\alpha_m|/S$ and the unnormalized state
\begin{equation}
\label{eq:gamma_bound}
   \ket{\psi_m}:=\ket{\psi_A} + \frac{\alpha_m}{p_m}\ket{m}
                =\ket{\psi_A}+S\sgn(\alpha_m)\ket{m},
   \qquad
   \Gamma:=\|\ket{\psi_m}\|=\sqrt{1-\epsilon^2+S^2},
\end{equation}
where $\sgn(\alpha)$ return the sign of $\alpha$. Let $\ket{\tilde{\psi}_m}:=\Gamma^{-1}\ket{\psi_m}$ be the normalized state. Then the deviation parameters satisfy:
\begin{equation}
  a:=\max_{m\in B}\left\|\ket{\tilde{\psi}_m}-\ket{\psi}\right\|
       \le (c+2)\epsilon+\bigO{\epsilon^2},
  \qquad
  b:=\left\|\sum_m p_m\ket{\tilde{\psi}_m}-\ket{\psi}\right\|
       = \bigO{\epsilon^2}.
\end{equation}
Consequently, the trace norm error of $\rho_{\mathrm{approx}} = p_m \ket{\tilde{\psi}_m}\bra{\tilde{\psi}_m}$ is $\|\rho_{\mathrm{approx}}-\ket{\psi}\bra{\psi}\|_1 = \bigO{\epsilon^2}$.
\end{theorem}

\begin{proof}
Using $\ket{\psi_m}=\ket{\psi_A}+\alpha_m/p_m\ket{m}$ and
$\ket{\psi}=\ket{\psi_A}+\alpha_m\ket{m}
           +\sum_{k\neq m}\alpha_k\ket{k}$ we obtain
\begin{align}
  \ket{\psi_m'}-\ket{\psi}
  &=\Gamma^{-1}\ket{\psi_m}-\ket{\psi}
   \notag\\
  &=\left(\Gamma^{-1}-1\right)\ket{\psi_A}
    +\left(\tfrac{\alpha_m}{p_m\Gamma}-\alpha_m\right)\ket{m}
    -\sum_{k\neq m}\alpha_k\ket{k}.        \label{eq:diff}
\end{align}

By \eqref{eq:gamma_bound}, $\Gamma=1+O(\epsilon^2)$, so $|\Gamma^{-1}-1|=O(\epsilon^2)$. Because $p_m=|\alpha_m|/S$, the middle term in Eq.~\eqref{eq:diff}
\begin{equation}
   \left|\tfrac{\alpha_m}{p_m\Gamma}-\alpha_m\right|
     \le \left|\tfrac{S}{\Gamma}\right|+ |\alpha_m|
     \le (c+1)\epsilon+O(\epsilon^2),
\end{equation}
where we used $|\alpha_m|\le\epsilon$ and~\eqref{eq:l1small}. The remaining $B$-components satisfy
$
  \left\|\sum_{k\neq m}\alpha_k\ket{k}\right\|
   =\sqrt{\epsilon^2-\alpha_m^2}\le\epsilon.
$

\noindent Combining the three parts gives
\begin{equation}
   \left\|\ket{\tilde{\psi}_m}-\ket{\psi}\right\|
     \le (c+2)\epsilon+O(\epsilon^2)
     \quad\forall m\in B,
\end{equation}
hence $a\le(c+2)\epsilon+O(\epsilon^2)$.

The unnormalized average equals the target state:
$\sum_m p_m\ket{\psi_m}=\ket{\psi}$.  Since $\Gamma$ is the
same for every $m$
\begin{equation}
   \left \|\sum_m p_m\ket{\psi_m'} - \ket{\psi}\right \|
     =\left \|(\Gamma^{-1}-1)\ket{\psi}\right \| \le \left \|\left(\frac{c^2-1}{2}\epsilon^2 + O(\epsilon^4)\right)\ket{\psi}\right \|
\end{equation}
so $b=\frac{c^2-1}{2}\epsilon^2 + O(\epsilon^4)$.

Applying the state-level mixing lemma yields
$
  \|\rho_{\mathrm{mix}}-\rho\|_1\le a^2+2b= (2c^2 + 4c +3)\epsilon^2 + \bigO{\epsilon^4} = \bigO{\epsilon^2}.
$
\end{proof}
This Theorem implies that we indeed achieve a quadratic speed up using the randomized scheme upon the assumption $l_1$-smallness condition is satisfied. In short, this assumption is justified if the coefficients with index in $B$ satisfy an exponential or a power-law exponent greater than 1. 

We first verify that the condition does not hold in general. Let $K:=|B|$ and ${\alpha}=(\alpha_i)_{i \in B}$. By Cauchy--Schwarz,
\begin{equation}
  S=\|{\alpha}\|_1  \le  \sqrt{K} \|{\alpha}\|_2
   = \sqrt{K} \epsilon.
  \label{eq:cs}
\end{equation}
Thus, a dimension-free constant $c$ in \eqref{eq:l1small} cannot be guaranteed from $\|{\alpha}\|_2=\epsilon$ alone unless the tail on $B$ is $\bigO{1}$ sparse.

In the following we explain when \eqref{eq:l1small} is satisfied under the decay assumption of the tail coefficients, which are common in practice. Let $|\alpha|_{(1)}\ge |\alpha|_{(2)}\ge \cdots$ denote the decreasing rearrangement of $\{|\alpha_i|\}_{i \in B}$, and write
\begin{equation}
S=\sum_{i \in B}|\alpha_i|=\sum_{j\ge 1} |\alpha|_{(j)},\qquad
\epsilon^2=\sum_{i \in B}\alpha_i^2=\sum_{j\ge 1} |\alpha|_{(j)}^2.
\end{equation}
We now connect the explicit decay models to the assumption $S\le c \epsilon$. Beginning by Geometric (exponential) decay. Assume
\begin{equation}
|\alpha|_{(j)} \le C r^{ j-1},\qquad 0<r<1. \label{eq:geom}
\end{equation}
Then
\begin{equation}
\frac{S}{\epsilon}
 \le 
\frac{\sum_{j\ge 1} Cr^{j-1}}{\Big(\sum_{j\ge 1} C^2r^{2(j-1)}\Big)^{1/2}}
=\frac{1 - r^K}{1-r}\sqrt{\frac{1-r^2}{1-r^{2K}}} \le \sqrt{\frac{1+r}{1-r}}
=:c(r),
\end{equation}
which is finite and independent of $|B|$. 

If we have a less strong power-law decay for the coefficients, assumption \eqref{eq:l1small} still holds if the decay power $r$ is greater than 1. 
Assume
\begin{equation}
|\alpha|_{(j)} \le C  j^{-r},\qquad r>\tfrac12. \label{eq:power}
\end{equation}
Then, in the infinite case, where $|B|\to\infty$,
\begin{equation}
\frac{S}{\epsilon}
 \sim 
\frac{\sum_{j\ge 1} j^{-r}}{\Big(\sum_{j\ge 1} j^{-2r}\Big)^{1/2}}
 = 
\begin{cases}
\dfrac{\zeta(r)}{\sqrt{\zeta(2r)}}, & r>1,\\ 
\text{diverges like } \log|B|, & r=1,\\
\text{diverges like } |B|^{ 1-r}, & \tfrac12<r<1,
\end{cases}
\end{equation}
where $\zeta(x) = \sum_{n=1}^{\infty} \frac{1}{n^x}$ is the Riemann zeta function.

\section{Configuration Interaction and Implementation Details}

\subsection{System Description and Mapping}
We validate our protocol using the Lithium Hydride (LiH) molecule, a standard benchmark in quantum chemistry. The molecule is modeled with a bond length of 1.6 \AA\ using the STO-3G minimal basis set. This basis yields 12 spin-orbitals (6 spatial orbitals).
The molecular electronic Hamiltonian in the second-quantization formalism is given by:
\begin{equation}
\hat{H} = \sum_{p,q} h_{pq} \hat{a}_p^\dagger \hat{a}_q + \frac{1}{2} \sum_{p,q,r,s} h_{pqrs} \hat{a}_p^\dagger \hat{a}_q^\dagger \hat{a}_s \hat{a}_r,
\end{equation}
where $\hat{a}_p^\dagger$ and $\hat{a}_p$ are fermionic creation and annihilation operators, and $h_{pq}$ and $h_{pqrs}$ represent one-electron and two-electron integrals, respectively. To process this on a quantum computer, we map the fermionic operators to qubits using the Jordan-Wigner transformation.

We compute the Full Configuration Interaction (FCI) ground state, $\ket{\Psi_{\mathrm{CI}}}$, by diagonalizing the Hamiltonian in the subspace of 4 electrons distributed across 12 spin-orbitals. The resulting state is a superposition of Slater determinants $\ket{\Phi_I}$:
\begin{equation}
    \ket{\Psi_{\mathrm{CI}}} = c_0 \ket{\Phi_0} + \sum_{I \neq 0} c_I \ket{\Phi_I},
\end{equation}
where $\ket{\Phi_0}$ is the Hartree-Fock reference state and $\ket{\Phi_I}$ represents excited configurations.

\subsection{Amplitude Hierarchy via Perturbation Theory}
The suitability of our randomized protocol for quantum chemistry stems from the natural hierarchy of the CI coefficients $c_I$. This hierarchy can be understood via Rayleigh--Schr\"odinger perturbation theory. Partition the Hamiltonian as $\hat{H} = \hat{H}_0 + \hat{V}$, where $\hat{H}_0$ is the M\o ller--Plesset zeroth-order Hamiltonian (sum of Fock operators).
To first order, the coefficients for excited determinants are:
\begin{equation}
    |c_I^{(1)}| = \left|-\frac{\bra{\Phi_I} \hat{V} \ket{\Phi_0}}{E_I^{(0)} - E_0^{(0)}}\right| \le \frac{\|\hat{V}\|}{\Delta E_{I0}} ,
\end{equation}
where $\Delta E_{I0} = E_I^{(0)} - E_0^{(0)}$ is the unperturbed energy gap.  
\begin{figure}
    \centering
    \includegraphics[width=0.5\linewidth]{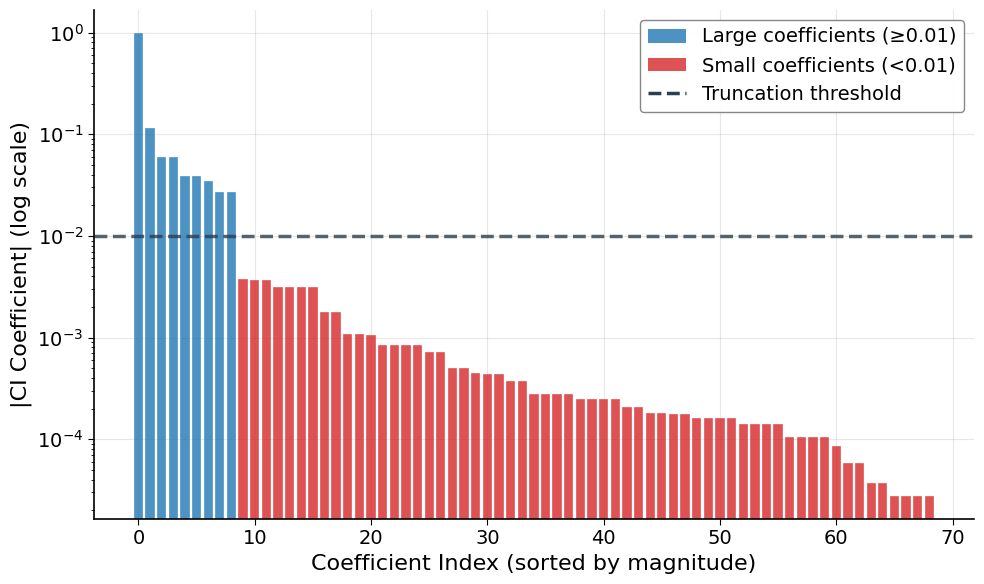}
    \caption{Coefficient distribution of the CI state for LiH molecule.}
    \label{fig:coefficientsdis}
\end{figure}
Even in higher-order perturbation theory or in full CI solutions, large denominator gaps $\Delta E_{I0}$ strongly suppress amplitudes for high-lying excited determinants. Consequently, the CI vector is naturally decaying: a few coefficients (HF and singles/doubles) carry the majority of the weight, while the "tail" of higher excitations decays rapidly. This structure ensures that the condition on $S$ (Eq.~\ref{eq:l1small}) is satisfied, maximizing the efficiency of our randomized truncation. We can see through Fig.~\ref{fig:coefficientsdis} that the majority of coefficients have small magnitudes, while there are in total 69 coefficients in the FCI state of LiH molecule, only 2 of them are greater than 0.1 and only 9 of them are greater than 0.01. 

\subsection{Randomized Implementation}
In our numerical implementation (Fig. 1 of main text), we:
1.  Compute the exact FCI vector $\mathbf{c} = \{c_I\}$ for LiH.
2.  Define a threshold $t$ and separate indices into $A$ and $B$.
3.  Calculate the normalization $\Gamma = \sqrt{\sum_{i \in A} |c_i|^2 + S^2}$.
4.  Construct the ensemble. For each $m \in B$, the circuit prepares a state proportional to $\sum_{i \in A} c_i \ket{x_i} + S \sgn(c_m) \ket{x_m}$.
    
Since the states in the ensemble differ only by the index and sign of the single amplified tail component, the compilation strategy (e.g., using Gleinig-Hoefler or similar sparse-state loaders) remains nearly identical for all $m$, simplifying control logic.

\section{Compilation and Resource Estimation}
\label{sec:compilation}

\subsection{Circuit Synthesis Methodology}
To rigorously quantify the resource savings, we generated explicit quantum circuits for every data point plotted in the main text.
We utilized a Python implementation of the state preparation algorithm proposed by Gleinig and Hoefler~\cite{gleinigEfficientAlgorithmSparse2021a}, which is asymptotically optimal for sparse state preparation. This algorithm takes a list of non-zero coefficients and their indices as input and outputs a quantum circuit composed of Clifford gates and multi-controlled SU(2) rotations.

For the numerical benchmarks:
\begin{itemize}
    \item \textbf{LiH Instance:} We synthesized circuits for truncation sizes (number of kept amplitudes) ranging from $K=1$ to $66$ for the randomized approach and from $K=1$ to $68$ for deterministic truncation.
    \item \textbf{Power-Law Instance:} We synthesized circuits for truncation sizes ranging from $K=1$ to around $700$ for both approaches.
\end{itemize}

For the randomized protocol, the output state is an ensemble $\left\{\ket{\tilde{\psi}_m}\right\}$. Since every state in this ensemble shares the exact same set of large coefficients $A$ and differs only by the index and sign of a single amplified tail coefficient $m$, the resulting circuit structures and complexities are nearly identical across the ensemble. Therefore, to estimate the resource cost for a given truncation level, we selected a single representative instance from the ensemble and performed the compilation and counting on that instance.

\subsection{Gate Counting and Decomposition}
The high-level circuits produced by the synthesis algorithm were decomposed into a standard fault-tolerant gate set (Clifford + $T$) to obtain the final counts.

\textbf{CNOT Counts:}
The multi-controlled SU(2) rotations were decomposed into CNOT gates and three single-qubit rotations using standard circuit identities. We recorded the exact total number of CNOT gates in the fully decomposed circuit.

\textbf{T-Gate Counts:}
We assume a fault-tolerant setting where arbitrary single-qubit rotations ($R_x, R_y, R_z$) are approximated using sequences of $H$ and $T$ gates. We employed the cost model for ancilla-free $HT$-decomposition~\cite{rossOptimalAncillafreeClifford+T2016}, which provides an upper bound on the $T$-count for synthesizing a rotation with error $\delta$:
\begin{equation}
    N_T \approx 3 \log_2\left(\frac{1}{\delta}\right).
\end{equation}
A critical factor in this estimation is the choice of the synthesis precision $\delta$. To ensure that the gate synthesis error does not overwhelm the inherent state preparation error, we set the target precision for each rotation gate adaptively: $\delta = \frac{\theta_{\min}}{m},$ where $m$ is the total number of single-qubit rotation gates in the circuit, and $\theta_{\min}$ is the minimum rotation angle (in radians) appearing in the circuit.

This adaptive precision criterion highlights the mechanism of our $T$-gate reduction. In the deterministic truncation, retaining small amplitudes without amplification results in very small rotation angles $\theta_{\min}$, necessitating extremely high precision $\delta$ (and thus many $T$ gates) to resolve them. In the randomized approach, small amplitudes are amplified, effectively increasing $\theta_{\min}$, which relaxes the required precision $\delta$ and significantly lowers the $T$-count.

\end{document}

%% file: preamble.tex
\usepackage{amsmath,amsthm,amssymb,epsfig,graphicx,mathrsfs,amsfonts,dsfont,bbm}
\usepackage{dcolumn}
\usepackage{bm}
\usepackage{tikz}
\usepackage{graphicx}
\usepackage{xcolor}
\usepackage[colorlinks=true,citecolor=magenta,urlcolor=cyan,linkcolor=black,filecolor=green]{hyperref}
\usepackage[capitalise]{cleveref}
\usepackage{enumitem}
\setlist[itemize]{leftmargin=20pt}
\setlist[enumerate]{leftmargin=20pt}
\usepackage{amsmath,amsfonts,amssymb}
\usepackage{braket}
\usepackage{physics}
\usepackage[linesnumbered,ruled,vlined]{algorithm2e}
\usepackage{algpseudocode}
\usepackage{multirow}
\usepackage{adjustbox}
\usepackage{float}
\usepackage{ragged2e} 
\usepackage[caption=false]{subfig}

\newtheorem{theorem}{Theorem}
\newtheorem{lemma}{Lemma}
\newtheorem{corollary}{Corollary}

\newcommand{\pow}{\textup{pow}}

\newcommand{\prlsection}[1]{\textit{#1.\textemdash}}

\newcommand{\vertiii}[1]{{\left\vert\kern-0.25ex\left\vert\kern-0.25ex\left\vert #1
		\right\vert\kern-0.25ex\right\vert\kern-0.25ex\right\vert}}
\newcommand{\Vertiii}[1]{{\vert\kern-0.25ex\vert\kern-0.25ex\vert #1
		\vert\kern-0.25ex\vert\kern-0.25ex\vert}}

\newcommand{\bigO}[1]{\mathcal{O}\left(#1\right)}

\newcommand{\sgn}{\text{sgn}}
\newcommand{\rand}{\text{rand}}
\newcommand{\Exp}{\text{exp}}